\documentclass[prd,tightenlines,twocolumn]{revtex4}
\usepackage{amsmath}
\usepackage{amssymb}
\usepackage{graphicx}
\usepackage{bm}
\usepackage{enumerate}
\usepackage{color}
\newcommand{\be}{\begin{equation}}
\newcommand{\ee}{\end{equation}}
\newcommand{\bea}{\begin{eqnarray}}
\newcommand{\eea}{\end{eqnarray}}
\newcommand{\ba}{\begin{eqnarray}}
\newcommand{\ea}{\end{eqnarray}}

\begin{document}

\title{Are there flux tubes in quark-gluon plasma?}
\author{ 
 Edward Shuryak }

\affiliation{Department of Physics and Astronomy, \\ Stony Brook University,\\
Stony Brook, NY 11794, USA}

%\date{\today}

\begin{abstract} 
We review both lattice evidences and theoretical arguments supporting the
existence of electric  flux tubes in quark-gluon plasma, above the deconfinement transition temperature $T>T_c$. At the end of this comment, we also point out certain 
questions which still needs to be answered.
 \end{abstract}

\maketitle

Let us start with two popular statements:

{\em (i) Existence of flux tubes between two fundamental charges in QCD-like gauge theories is among the most direct
manifestations of the confinement
phenomenon.

(ii) Confinement  is well described by the ``dual superconductor" model \cite{Mandelstam:1974pi,tHooft:1977nqb} , relating it to
known properties of superconductors via electric-magnetic duality.}

In this note we discuss both of them, arguing that
they are only $partially$ correct. In fact we are going to argue that flux tubes
do exist even above the deconfinement transition temperature, and that
there are much better analogies to confinement than
superconductors. Most facts and considerations needed to understand the phenomena discussed has in  fact been
in literature for some time. The purpose of these comments are simply to remind them, ``connecting the dots" once again, since these questions continue to be asked at the meetings. We will also point out  certain aspects of the phenomena 
which still need to be clarified.

\section{Flux tubes on the lattice, at zero $T$ and near $T_c$}

Lattice gauge theory simulations have addressed the confinement issue from their
beginning, and by now there are many works which studied the
electric flux tubes between static charges. Most of those are done at zero/low $T$.
The documented well the profile of the electric field and the  magnetic current
``coiling" around it. 

\begin{figure}[h!]
\begin{center}
\includegraphics[width=7cm]{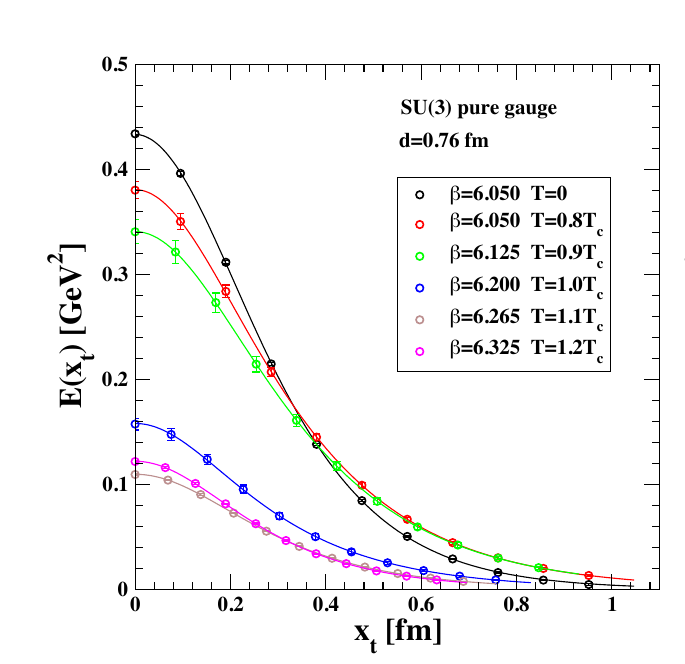}
\caption{ The longitudinal electric field as a function of $transverse$ coordinate is measured,
for a number of temperatures, for pure gauge $SU(3)$ theory, from  \cite{Cea:2017bsa}.}
\label{fig_profile_T}
\end{center}
\end{figure}

The ``dual superconductor" analogy leads to a comparison with Ginzburg-Landau theory
 (also called  ``the dual Higgs model") and
 good agreement with it has been found. These results 
 are well known for two decades, see e.g.  the review \cite{Bali:1998de}.  
 
However, recent lattice studies \cite{Cea:2017bsa} exploring a near-deconfinement range of temperatures
have found that a tube-like profile of the electric field persists even above the critical temperature $T_c$, to
at least  $1.5 T_c$. One of the plots from this work is reproduced in Fig.\ref{fig_profile_T}.
It corresponds to pure gauge $SU(3)$ theory, which has
the first order transition, seen as a jump in the field strength. Note however that at $T>T_c$
the shape of the electric field transverse profile remains about the same, while the   width is $decreasing$ with $T$, making it even more tube-like, rather than expected near-spherical Coulomb behavior. (Note also that the length of the flux tube remains constant, $0.76\, fm$ for this plot.)

Such behavior clearly contradicts the ``dual superconductor" model:
in superconductors, the flux tubes are only observed in the 
superconducting phase.
Why do we observe flux tubes in the ``normal" phase, and do we really have any contradictions with theory here?

\section{Does the $T_c$ indeed represent the monopole condensation temperature?} 

Let us start by critically examining the very notion of the deconfinement transition itself,
focusing on whether it is indeed is the transition between the super and the normal phases. 

The $T_c$ itself is defined from thermodynamical quantities, and for pure gauge theories (we will only consider in this note)
its definition has no ambiguities.  

At this point it is worth reminding that in fact the electric-magnetic duality relates QCD not to the BCS superconductors, but rather
to  Bose-Einstein condensation (BEC) of bosons, the magnetic monopoles.
 Multiple lattice studies did confirmed that  $T_c$  does coincide with BEC of monopoles.

One such study I was involved in   \cite{D'Alessandro:2010xg} has calculated the probability
of the so called Bose (or rather Feynman's) clusters, a set of $k$ monopoles  
 interchanging their locations over the Matsubara time period. Its dependence on $k$
 leads to the definition of the  effective chemical potential, which is shown to vanish
 exactly at $T=T_c$. This means that monopoles do behave as any other bosons, and they indeed undergo
Bose-Einstein condensation at exactly $T=T_c$.

Earlier studies  by  Di Giacomo
and collaborators in Pisa group over the years, see e.g.  \cite{Bonati:2011jv}, were based
on the idea to construct (highly non-local) order parameter for monopole BEC. 
 It calculates the temperature dependence of the expectation value of the operator, effectively inserting/annihilating a monopole, and indeed finds a jump exactly at $T_c$.

In summary, it has been shown beyond a reasonable doubt that $T_c$ does indeed separate the ``super" and ``normal" phases.

\section{Constructing the flux tubes in the ``normal" phase}
Since electric-magnetic duality relates QCD not to the BCS superconductors, but rather
to BEC, let us at this point emphasize a
 significant difference between them:  the  $uncondenced$ bosons are also
present in the system, both above and even below $T_c$, while the BCS Cooper pairs of  superconductor exist at $T<T_c$ only.

The first construction of the flux tube in the normal phase has been made by
Liao and myself \cite{Liao:2007mj}. The key point is  that it   does not require supercurrents.
Indeed, various flux tubes are found in plasmas: e.g. one can even see them in solar corona in an average telescope.
What is needed for flux tube formation is in fact  the presence of {\em dual plasma}, a medium including moving magnetic charges. 

\begin{figure}[h!]
\begin{center}
\includegraphics[width=6cm]{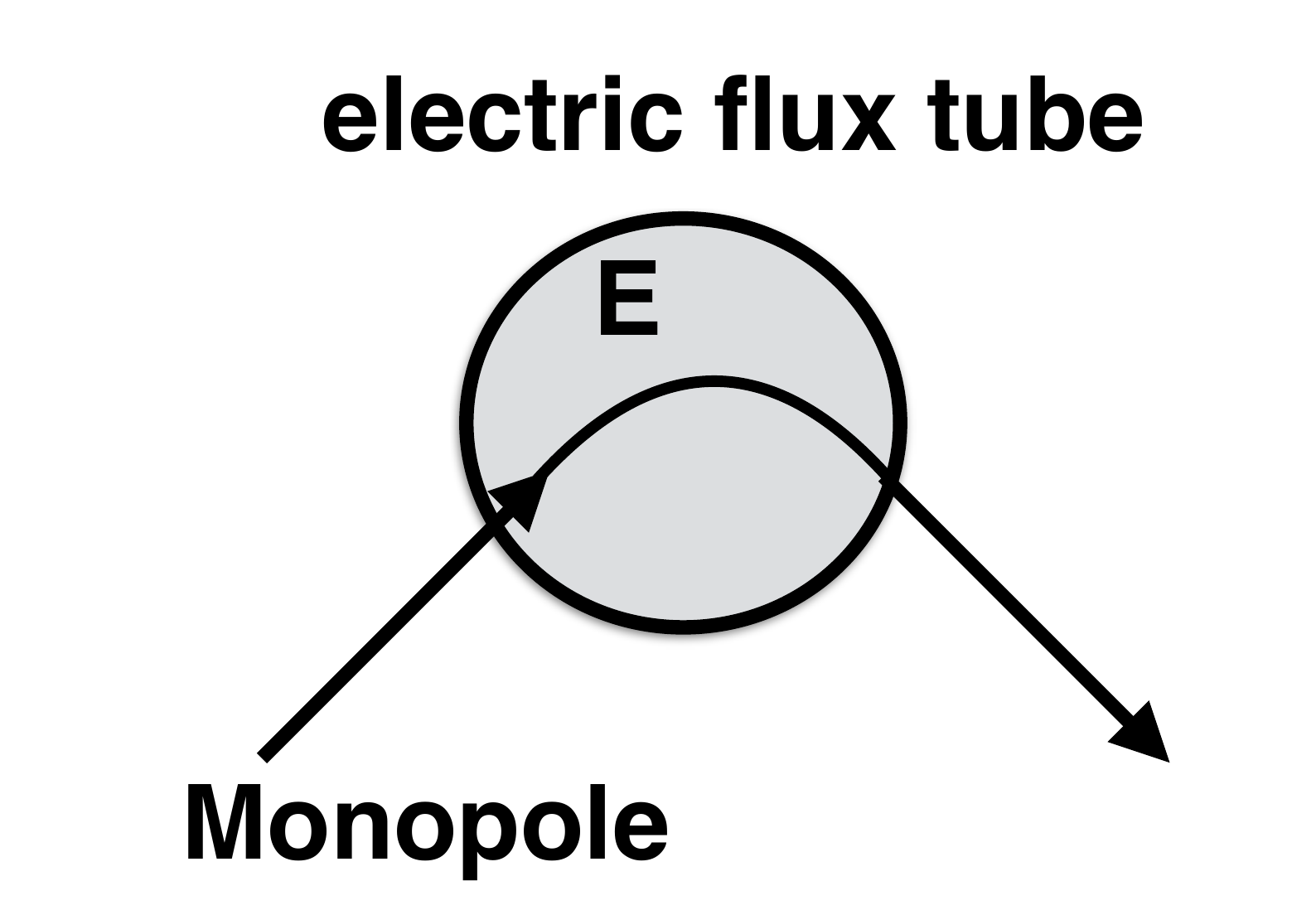}
\caption{A sketch of a monopole traversing the electric flux tube}
\label{fig_flux_tube_mono}
\end{center}
\end{figure}

Their scattering 
on the electric flux tube schematically shown in Fig.\ref{fig_flux_tube_mono} does not change
the monopole energy but changes direction of its momentum, thus
creating a force on the flux tube. If it is strong enough
 able to confine the electric field, a flux tube solution can be constructed. 
 
 For further 
 details see the original paper   \cite{Liao:2007mj}. Let us only comment that (i)
 the ``uncondenced' monopoles exert a $larger$ force than those in the condensate, as their momenta are larger; and (ii)  
   it has in fact been $predicted$
 there that the highest $T$ at which such solution may exist is about $1.5T_c$.   

 \begin{figure}[h]
\begin{center}
\includegraphics[width=5.5cm]{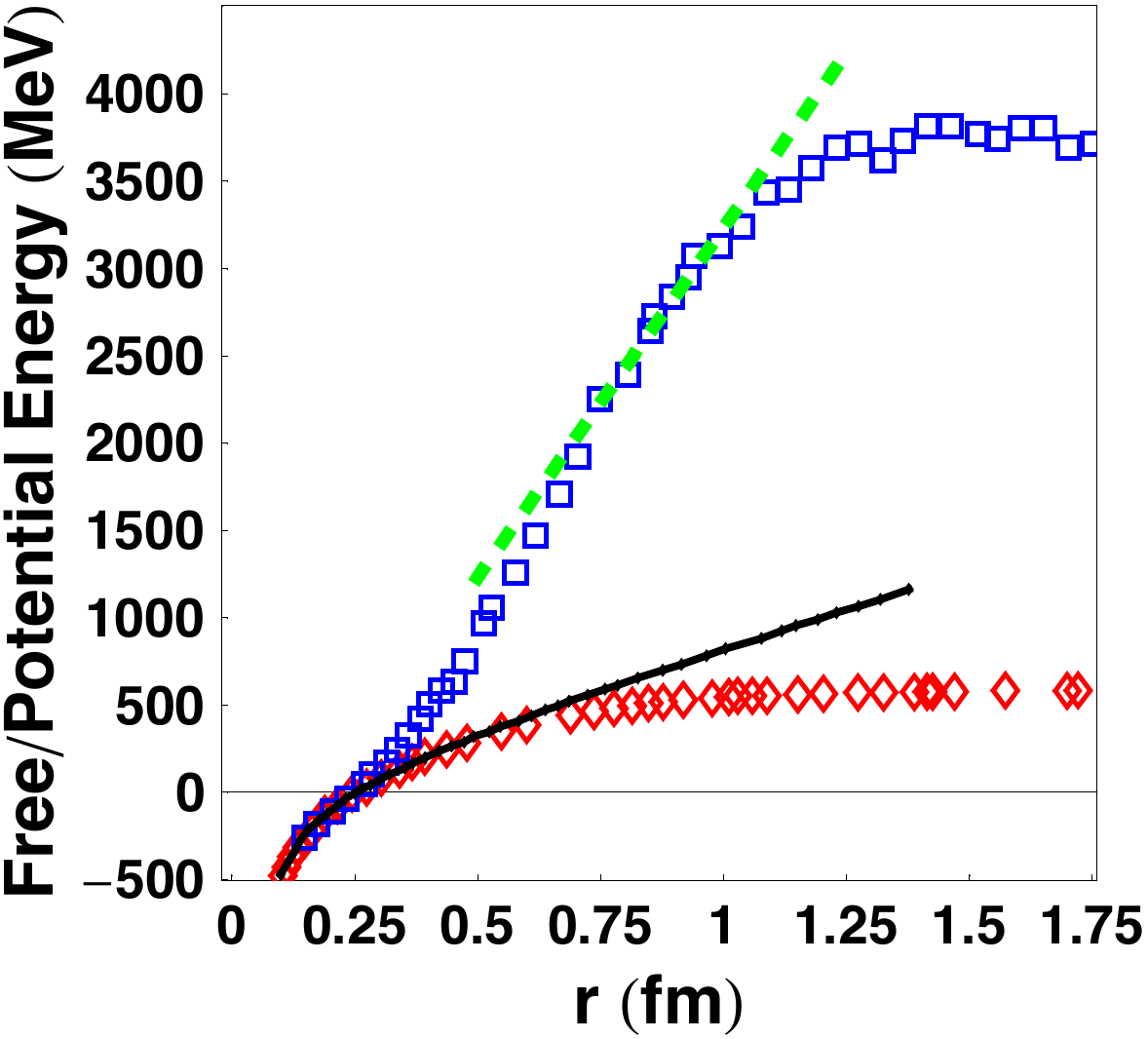}
\includegraphics[width=6cm]{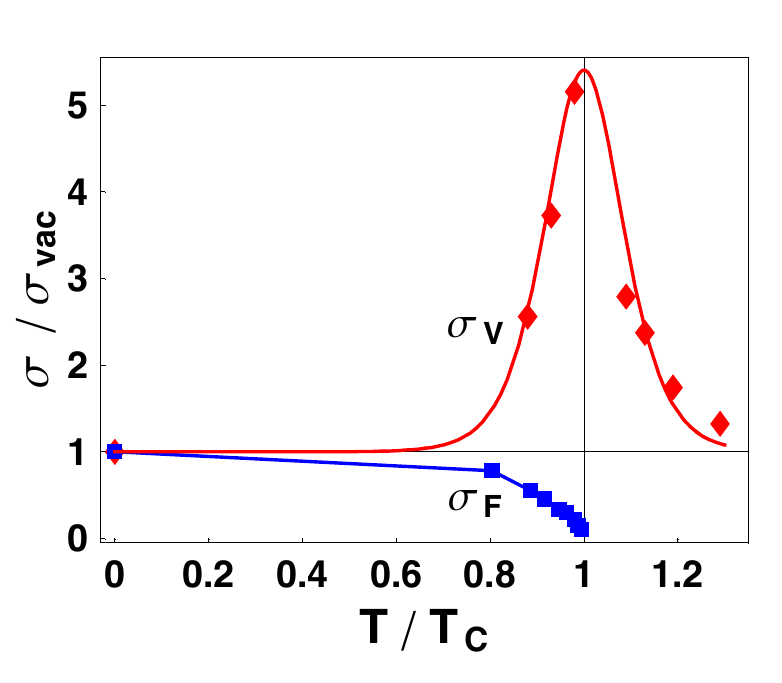}
\caption{Left: Free (red rhombs) energy $F(r)$ and potential (blue squares) energy $V(r)$,
at $T_c$, compared to the zero temperature potential (black line). 
Right: Effective string tension for the free 
 and the internal energy, from \cite{Kaczmarek:2005gi}.} 
%Right: thermal monopole density $n_M /T^3$ normalized to $T^3$.  The two curves connecting box symbols are
%for $\alpha_E= 0.5$ (upper, red) and 1(lower, blue) respectively, while the two green curves connecting diamond symbols showing SU(2)(lower) lattice data and their SU(3) extrapolation(upper) for $T >1.3T_c $. }
\label{fig_tension_F_U}
\end{center}
\end{figure}
 
\section{Two static potentials and the flux tube entropy issue}

So far there were no subtleties involved. But let us  asks now the following question: Provided  these flux tubes at $T>T_c$
 carry some tension (energy-per-length), does it still imply existence of a linear potential between quarks, up to  $T=1.5T_c$? And if they do,  would  it imply that, in a sense, confinement remains  enforced  there?

In order to have proper perspective on  what is going on, let us look back at lattice studies of the static quark (fundamental) potentials, e.g. 
 \cite{Kaczmarek:2005gi}. The key point is that 
 there are $two$ kinds of the potentials. At finite temperatures the natural quantity to calculate, for the observed flux tubes
between static charges, is the {\em free energy}. It can be written as
 \be F(r) =V(r)- T S(r), \,\,\, S(r)={\partial F \over \partial r}   \ee 
where $S(r)$ is the $entropy$ associated with a pair of static quarks. 
Since it can be calculated from the free energy itself, as indicated in the r.h.s., one can 
subtract it and plot also the {\em potential energy} $V(r)$. The derivatives over $r$ are known as  the {\em string tensions}.

These lattice calculations have shown that in certain range of $r$ the tension is
constant (the tension is approximately $r$-independent).  Two resulting tensions, 
shown in Fig.\ref{fig_tension_F_U}(left) have very different temperature dependence.
The tension of the free energy shows the expected behavior: $\sigma_F(T)$ vanishes as
$T\rightarrow T_c$.  But the  tension of the potential energy $\sigma_V(T)$
shows drastically different behavior, with large $maximum$ at $T_c$, and non-zero value above it. This unexpected behavior was hidden in $\sigma_F(T)$, studied in many previous works, because in it a large energy and a large entropy cancel each other. 
  
So, everything would be consistent, provided these novel flux tubes at $T>T_c$ do 
indeed carry the $potential$ energy only, but no $free$ energy tension $\sigma_V\neq 0, \sigma_F=0$.  In other words, we suggest that the potential energy detected (via electric field squared) in \cite{Cea:2017bsa} must be canceled by the  entropy associated with
it, and no actual force between charges would be present in equilibrium.  This conjecture can and should be checked. 

Similar comment applies also to the theoretical calculation of the potential: 
 in \cite{Liao:2007mj} only the mechanical stability of the tube solution was derived.
 The entropy associated with the flux tube still remains to be calculated. 

As a parting comment, while this conjecture sounds like the well known idea of Hagedron string
transition, it cannot be exactly that. Indeed, this idea is known to suggest that at $T>T_c$ 
string gets to be infinitely long. If so, the tube completely delocalizes, and there would be a
Coulomb field rather than what was observed by  \cite{Cea:2017bsa}. The entropy
in question is perhaps related to monopoles  bound to the tube rather than its multiple shapes. 
Also a Hagedorn transition seems to be at odds with the tension increase and size decrease
as a function of $T$ observed.

(Finally, let us for  clarity mention that we only discuss in this note static potentials in thermal equilibrium. We do not discuss potentials in quarkonia, in which quarks are not standing but moving.  This problem is associated with certain time scales, inducing deviation from equilibrium and possible dissipation. It would therefore require a completely separate 
discussion.)


\begin{thebibliography}{99} 



%\cite{tHooft:1977nqb}
\bibitem{tHooft:1977nqb} 
  G.~'t Hooft,
  ``On the Phase Transition Towards Permanent Quark Confinement,''
  Nucl.\ Phys.\ B {\bf 138}, 1 (1978).
  doi:10.1016/0550-3213(78)90153-0
  %%CITATION = doi:10.1016/0550-3213(78)90153-0;%%
  %1436 citations counted in INSPIRE as of 25 Jun 2018	*** Not Found with lookup: '035__a:'tHooft:1977hy'

%\cite{Mandelstam:1974pi}
\bibitem{Mandelstam:1974pi} 
  S.~Mandelstam,
  ``Vortices and Quark Confinement in Nonabelian Gauge Theories,''
  Phys.\ Rept.\  {\bf 23}, 245 (1976).
  doi:10.1016/0370-1573(76)90043-0
  %%CITATION = doi:10.1016/0370-1573(76)90043-0;%%
  %1125 citations counted in INSPIRE as of 25 Jun 2018


%\cite{Bali:1998de}
\bibitem{Bali:1998de} 
  G.~S.~Bali,
  ``The Mechanism of quark confinement,''
  hep-ph/9809351.
  %%CITATION = HEP-PH/9809351;%%
  %77 citations counted in INSPIRE as of 25 Jun 2018

%\cite{Cea:2017bsa}
\bibitem{Cea:2017bsa} 
  P.~Cea, L.~Cosmai, F.~Cuteri and A.~Papa,
  %``QCD flux tubes across the deconfinement phase transition,''
  EPJ Web Conf.\  {\bf 175}, 12006 (2018)
  doi:10.1051/epjconf/201817512006
  [arXiv:1710.01963 [hep-lat]].
  %%CITATION = doi:10.1051/epjconf/201817512006;%%
  %1 citations counted in INSPIRE as of 25 Jun 2018


%\cite{D'Alessandro:2010xg}
\bibitem{D'Alessandro:2010xg} 
  A.~D'Alessandro, M.~D'Elia and E.~V.~Shuryak,
  ``Thermal Monopole Condensation and Confinement in finite temperature Yang-Mills Theories,''
  Phys.\ Rev.\ D {\bf 81}, 094501 (2010)
  doi:10.1103/PhysRevD.81.094501
  [arXiv:1002.4161 [hep-lat]].
  %%CITATION = doi:10.1103/PhysRevD.81.094501;%%
  %47 citations counted in INSPIRE as of 25 Jun 2018


%\cite{Bonati:2011jv}
\bibitem{Bonati:2011jv} 
  C.~Bonati, G.~Cossu, M.~D'Elia and A.~Di Giacomo,
  ``The disorder parameter of dual superconductivity in QCD revisited,''
  Phys.\ Rev.\ D {\bf 85}, 065001 (2012)
  doi:10.1103/PhysRevD.85.065001
  [arXiv:1111.1541 [hep-lat]].
  %%CITATION = doi:10.1103/PhysRevD.85.065001;%%
  %12 citations counted in INSPIRE as of 25 Jun 2018


%\cite{Liao:2007mj}
\bibitem{Liao:2007mj} 
  J.~Liao and E.~Shuryak,
  ``Electric Flux Tube in Magnetic Plasma,''
  Phys.\ Rev.\ C {\bf 77}, 064905 (2008)
  doi:10.1103/PhysRevC.77.064905
  [arXiv:0706.4465 [hep-ph]].
  %%CITATION = doi:10.1103/PhysRevC.77.064905;%%
  %46 citations counted in INSPIRE as of 25 Jun 2018


%\cite{Kaczmarek:2005gi}
\bibitem{Kaczmarek:2005gi} 
  O.~Kaczmarek and F.~Zantow,
  ``Static quark anti-quark interactions at zero and finite temperature QCD. II. Quark anti-quark internal energy and entropy,''
  hep-lat/0506019.
  %%CITATION = HEP-LAT/0506019;%%
  %84 citations counted in INSPIRE as of 25 Jun 2018

%
%
%%\cite{Cea:2017bsa}
%\bibitem{Cea:2017bsa} 
% P.~Cea, L.~Cosmai, F.~Cuteri and A.~Papa,
%  ``QCD flux tubes across the deconfinement phase transition,''
%  EPJ Web Conf.\  {\bf 175}, 12006 (2018)
%  doi:10.1051/epjconf/201817512006
%  [arXiv:1710.01963 [hep-lat]].
%
%%\cite{Liao:2007mj}
%\bibitem{Liao:2007mj} 
%  J.~Liao and E.~Shuryak,
%  ``Electric Flux Tube in Magnetic Plasma,''
%  Phys.\ Rev.\ C {\bf 77}, 064905 (2008)
%  doi:10.1103/PhysRevC.77.064905
%  [arXiv:0706.4465 [hep-ph]].
%  %%CITATION = doi:10.1103/PhysRevC.77.064905;%%
%  %46 citations counted in INSPIRE as of 24 Jun 2018

\end{thebibliography}
\end{document}